\documentclass[aps,prb,twocolumn,superscriptaddress,showpacs,amsmath,amssymb]{revtex4}
\usepackage{graphicx}
\usepackage[floatfix]{epsfig}
\usepackage{dcolumn}
\usepackage{bm}

\def\tc{T$_c$ }
\def\sc{superconducting }

\def\beq{\begin{equation}}
\def\eeq{\end{equation}}
\def\eqref#1{Eq.~(\ref{#1}) }
\def\de{\partial}

\def\check#1{{\bf #1}}

\begin{document}


\title{Phase-fluctuation induced reduction of the kinetic energy 
at the superconducting transition
}


\author{T.~Eckl}
\email{eckl@physik.uni-wuerzburg.de}
\affiliation{Institut f\"ur Theoretische Physik und Astrophysik,
Universit\"at W\"urzburg, Am Hubland, D-97074 W\"urzburg, Germany}

\author{W.~Hanke}
\email{hanke@physik.uni-wuerzburg.de}
\affiliation{Institut f\"ur Theoretische Physik und Astrophysik,
Universit\"at W\"urzburg, Am Hubland, D-97074 W\"urzburg, Germany}

\author{E.~Arrigoni}
\email{arrigoni@physik.uni-wuerzburg.de}
\affiliation{Institut f\"ur Theoretische Physik und Astrophysik,
Universit\"at W\"urzburg, Am Hubland, D-97074 W\"urzburg, Germany}
\affiliation{Department of Physics, University of California, Los Angeles, CA 90095 USA}

\date{\today}

\begin{abstract}
Recent reflectivity measurements indicated a possible \emph{violation} of the in-plane optical integral in
the underdoped high-$T_c$ compound $Bi_2Sr_2CaCu_2O_{8+\delta}$ up to 
frequencies much higher than expected by standard BCS theory.
The sum rule violation may be related to a loss of in-plane kinetic energy
at the superconducting transition.
Here, we show that a model based on phase fluctuations of the superconducting
order parameter introduces a change of the in-plane kinetic energy at $T_c$.
The change is due to a transition from a phase-incoherent Cooper-pair motion
in the pseudogap regime above $T_c$ to a phase-coherent motion at $T_c$.
\end{abstract}


\maketitle

\section{Introduction}

The key idea of the phase-fluctuation scenario in the high-$T_c$ superconductors is the notion
that the pseudogap observed in a wide variety of experiments arises from phase fluctuations
of the superconducting gap \cite{em.ki.95,ra.tr.92,fr.mi.98,kw.do.99,herb.02,fr.te.01,ec.sc.02}.
In this scenario, below a mean-field temperature scale $T_c^{MF}$, a $d_{x^2-y^2}$-wave gap amplitude
is assumed to develop. However, the superconducting transition is suppressed to a considerably lower
transition temperature $T_c$ by phase fluctuations \cite{em.ki.95,ec.sc.02}. In the intermediate
temperature regime between $T_c^{MF}$ and $T_c$, phase fluctuations
of the superconducting order parameter give rise to the pseudogap phenomena.
Recently, we have shown that indeed a two-dimensional BCS-like Hamiltonian with a $d_{x^2-y^2}$-wave gap
and phase fluctuations, which were treated by a Monte-Carlo simulation of an $XY$ model, yields results
which compare very well with scanning tunneling measurements over a wide temperature range \cite{ec.sc.02,ku.fi.01}.
Thus, they support the phase-fluctuation scenario for the pseudogap.

There is also increasing evidence from a number of recent experiments for the
relevance of phase fluctuations such as the Corson et al.~measurements \cite{co.ma.99}
of the high-frequency conductivity that track the phase correlation time.
$XY$ vortices are probably responsible for the large Nernst effect \cite{wa.xu.00,wa.on.02}.
The evolution of $T_c$ with electron irradiation found very recently \cite{ru.al.03u} also emphasizes
the importance of phase fluctuations. In this paper, we argue that phase fluctuations should have yet
another rather unexpected consequence: they induce a reduction of the kinetic energy at the 
\sc transition. This reduction is due to a transition from a ``disordered'', i.~e.~phase-incoherent
Cooper-pair motion in the pseudogap regime above $T_c$ to an ``ordered'', i.~e.~phase-coherent
motion at $T_c$. Comparison of our results, based on the BCS phase-fluctuation model,
with optical experiments \cite{mo.pr.02,sa.lo.01u} support this idea.

In ordinary BCS superconductors  the optical conductivity is suppressed
at frequencies within a range of about twice the \sc gap. The corresponding
low-frequency spectral weight 
$W_{low}$ is 
 transfered to the zero-frequency 
delta peak $W_D$ \cite{tinkham}, associated with the dissipationless
transport (and the superfluid weight $D$) in the \sc state.
This is the Glover-Ferrell-Tinkham (GFT) sum rule. On the other hand,
the {\it total} frequency integral of the optical conductivity 
is conserved, when decreasing the temperature across the \sc transition,
due to the f-sum rule~\cite{tinkham}, i.~e.~$W_{tot}^{sc} = W_{tot}^n$.

However, recent measurements of the in-plane optical conductivity \cite{mo.pr.02,sa.lo.01u}
 have indicated a \emph{violation} of the GFT optical sum rule for frequencies up to $2 eV$
in underdoped $Bi_2Sr_2CaCu_2O_{8+\delta}$ (Bi2212). 
By entering the superconducting state, not only spectral weight $W_{low}$ 
from the microwave and far-infrared, but also from the visible optical spectrum i.~e.~high-frequency 
spectral weight $W_{high}$
contributes to the superfluid condensate $W_D$. 
That is, in contrast to ordinary BCS-superconductors, a ``\emph{color} change'' is introduced
at the superconducting transition.
The interpretation of this unusual result may require the inclusion of local-field effects and other (such as excitonic)
many-body effects. They are known to play a crucial role already in weakly-correlated systems (such as semiconductors),
and introduce a shift of order of the Coulomb correlation energy between single-particle and
two-particle, i.~e.~optical excitations \cite{ha.sh.80}. Therefore, they may partly account for the 
``high-energy'' features observed in $\sigma(\omega)$.
On the other hand, within a tight-binding one-band model,
the anomalously large energy scale, which contributes to the superfluid weight, and the
corresponding \emph{color} change can be attributed to a \emph{reduction of kinetic energy}  
\cite{hirs.02} at the superconducting transition.
This is rather surprising, since one would expect that in a
 conventional (BCS) pairing process, it is
the potential energy which is reduced at the expense of the kinetic energy,
with the latter being  
increased due to particle-hole mixing.

The full optical integral, when integrated over all frequencies and \emph{energy bands}, is
proportional to the carrier density ($n$) over the bare mass ($m$) 
\begin{equation}
W_{tot}\equiv W_{low}+W_D+W_{high}=\int_0^{\infty} Re\; \sigma_{xx}(\omega)\; d \omega = \frac{n e^2}{2 m},
\end{equation}
and, thus, is conserved.
When the optical integral is restricted over a finite (low) range of
frequencies $\Omega$, in the HTSC typically of the order
of eV, one may consider the weight $W_{low}+W_D$ as being essentially due to a single band around
the Fermi energy, i.~e.
\begin{equation}
W_{low}+W_D = \int_0^{\infty} Re\; \Tilde{\sigma}_{xx}(\omega)\; d \omega = (\pi e^2 a^2 / 2 \hbar^2 V) E_{K},
\end{equation}
where $\Tilde{\sigma}$ is the single-band conductivity, $a$ the lattice constant and $V$ the
unit cell volume.
With this single-band assumption,  the frequency  integral 
of the optical conductivity is proportional 
to the  the inverse mass tensor ($\frac{\de^2 \epsilon_k}{\de k_x^2}$,
$\hat x$ being the direction in which the conductivity is measured)
weighted with the momentum distribution $n_k$ \cite{no.pe.02u,kubo.57}:
\begin{equation}
 E_{K} = (2/a^2 N) \sum_k \frac{\de^2 \epsilon_k}{\de k_x^2} n_k,
\end{equation}
with $N$ the number of k points.
This quantity depends upon the bare single-particle band structure $\epsilon_k$, being
 proportional to minus the kinetic energy $E_K=-E_{kin}$ for
a (nearest-neighbor) tight-binding (TB) model, while for free electrons it is a constant given by the
electron density divided by the effective mass.

\section{Phase Fluctuation Scenario for Kinetic Energy Reduction}

In this paper, we propose as a mechanism for a kinetic-energy reduction phase-fluctuations.
That is, in order to have condensation into
the superconducting state,
one needs,
 in addition to the binding of charge carriers into Cooper pairs,
long-range phase coherence among the pairs.
Since superconductors with low superconducting carrier density
(such as the organic and underdoped high-$T_c$ superconductors) 
are characterized by a relatively small
phase \emph{stiffness}, this implies a significantly larger role for
phase fluctuations, than in conventional superconductors \cite{em.ki.95,ca.ki.99,klei.00}. As a
consequence, in these materials  the 
transition to the superconducting state does not display a typical
mean-field (BCS) behavior,  
and phase fluctuations, both classical and quantum, 
may have a significant influence on low temperature properties. When
coherence is lost due to  
thermal fluctuations of the phase at and above the transition temperature $T_c$,
pairing remains, together with short-range phase correlations. 
These phase fluctuations can cause
the pseudogap phenomena observed e.~g.~in tunneling experiments \cite{re.re.p.98,mi.za.99,ku.fi.01,ec.sc.02}
in the underdoped HTSC.

\begin{figure}[t]
\begin{center}
\epsfig{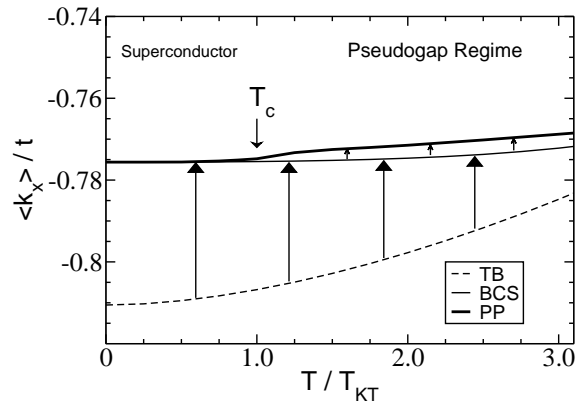}
\end{center}
\caption[]{Kinetic energy per bond $\langle k_x \rangle$ as a function of temperature for the non-interacting tight-binding 
electrons (TB), the BCS solution (BCS), and our phase-fluctuation model (PP) for $\mu = 0 $ ($\langle n \rangle = 1$).
The large vertical arrows indicate the increase in kinetic energy upon pairing relative to the free
tight-binding model, and the small arrows indicate the additional increase due to  
phase fluctuations. This additional \emph{phase-fluctuation energy} rapidly vanishes
near $T_c \equiv T_{KT}$, which causes the significant change in the optical integral
upon entering the superconducting state at $T_{KT}=0.1t$.
Note that the thick line follows the actual kinetic energy encountered in our
model, when going from the pseudogap to the superconducting regime.}
\label{ekin}
\end{figure}

\begin{figure}[t]
\begin{center}
\epsfig{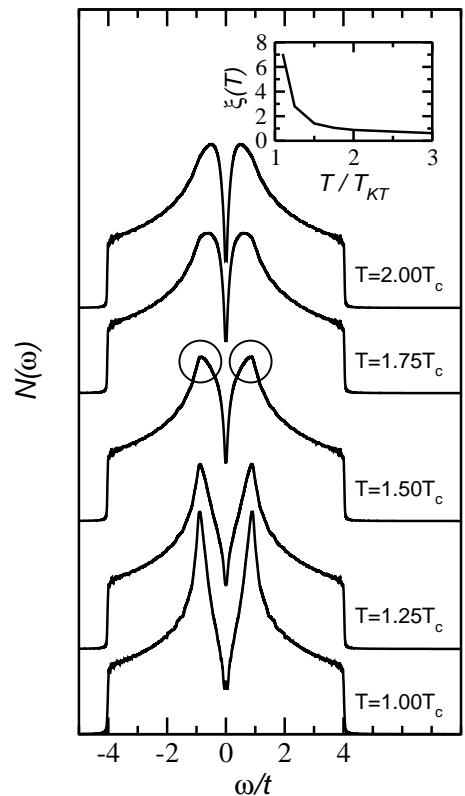}
\end{center}
\caption[]{Single particle density of states $N(\omega)$ for different temperatures
$T$ for a $32 \times 32 $ lattice. Coherence peaks develop (marked with circles for $T=1.5 T_c$) as $T$ approaches $T_c\equiv T_{KT}$
in exactly the same temperature regime ($T_c<T<1.5T_c$) where in Fig.~\ref{ekin} the kinetic energy reduction
occurs. The inset shows the corresponding temperature dependence of
the phase correlation length $\xi(T)$.}
\label{n_omega}
\end{figure}

We show here that indeed phase fluctuations
contribute to a significant reduction of  
the in-plane kinetic energy
upon entering into the \sc phase below $T_c$,
with a magnitude comparable to recent experimental results.
The physical reason for this kinetic energy lowering is
 that, 
 due to phase fluctuations and to the associated
 incoherent motion of Cooper
 pairs (cf. Fig.~\ref{ekin}),
the pseudogap region has a higher kinetic energy than the
 simple BCS mean-field state.
When long-ranged phase coherence finally
develops at \tc, 
the Cooper-pair motion becomes phase {\it coherent}
and the kinetic energy decreases. 
The onset of the coherent motion
can be seen, for example, from the
development of coherence peaks in the tunneling spectrum of BiSrCaCuO
compounds (see e.~g.~\cite{ku.fi.01,ec.sc.02} and Fig.~\ref{n_omega}).
The initial
\emph{cost} of kinetic energy, 
which is needed for pairing,
 is payed
 at a mean-field temperature
$T_c^{MF}$
considerably higher than $T_c$.
Therefore, the reduction of kinetic energy observed
experimentally~\cite{mo.pr.02,sa.lo.01u}  can
be attributed to a transition from a
phase-disordered pseudogap to a phase-ordered superconducting state.
We stress that this effect is independent of the particular mechanism
leading to pair formation, as long as the superconductor considered is characterized by
a small phase stiffness \cite{kivelson}.

Our starting Hamiltonian is of a simple BCS form given by
\begin{equation}
H=K -\frac{1}{4}
\sum_{i\,\delta}(\Delta_{i\,\delta}\langle\Delta_{i\,\delta}^\dagger\rangle 
+\Delta_{i\,\delta}^\dagger\langle\Delta_{i\,\delta}\rangle),
\label{one}
\end{equation}
with the nearest-neighbor hopping term
\begin{equation}
K=-t \sum_{\langle i\,j\rangle,\sigma}(c^\dagger_{i\,\sigma}c_{j\,\sigma}+c^\dagger_{j\,\sigma}c_{i\,\sigma}),
\label{oneprime}
\end{equation}
where $c^\dagger_{i\,\sigma}$ creates an electron of spin $\sigma$ on the $i^{\, th}$
site and $t$ denotes an effective nearest-neighbor hopping. The $\langle
i\,j\rangle$ sum is over nearest-neighbor sites of a 2D square lattice,
and, in the pairing term, $\delta$ connects $i$ to its nearest-neighbor sites.
The local $d$-wave gap,
\begin{equation}
\langle\Delta_{i\,\delta}^\dagger\rangle=
\frac{1}{\sqrt{2}}\langle c_{i\,\uparrow}^\dagger c_{i+\delta\,\downarrow}^\dagger
-c_{i\,\downarrow}^\dagger c_{i+\delta\,\uparrow}^\dagger\rangle=
\Delta\,e^{i \,\Phi_{i \delta}},\label{two1}
\end{equation}
is characterized by the \emph{fluctuating} phases
\begin{equation}
\Phi_{i \delta}=\left\{\begin{array}{l@{\quad \mathrm{for} \quad}l}
(\varphi_i + \varphi_{i+\delta})/2 & \text{$\delta$ in x-direction} \\
(\varphi_i + \varphi_{i+\delta})/2 +\pi & \text{$\delta$ in y-direction,} 
\end{array} \right. \label{two2}
\end{equation}
and by a spatially constant amplitude $\Delta$. 
We neglect the relative bond phase fluctuations between $\delta=\hat{x}$ and $\hat{y}$
as well as amplitude fluctuations. Thus, we consider only {\sl center of mass} pair phase
fluctuations, which are the relevant low-energy degrees of freedom, in a situation
in which the superfluid density is small, like in the underdoped cuprates.
In our two-dimensional
(2D) model $T_c$
corresponds to the
Kosterlitz-Thouless
transition temperature $T_{KT}$, where the phase correlation 
length $\xi$ diverges.

With this Hamiltonian it is straightforward to show, that the
optical sum rule yields \cite{sc.wh.93,kubo.57,delta},
\begin{equation}
\check{}
\int_0^{\infty} Re\; \Tilde{\sigma}_{xx}(\omega)\; d \omega = -e ^2 \pi \langle k_x \rangle / 2,
\end{equation}
in units where $\hbar=c=1$,
with $\langle k_x \rangle$ being the expectation value of the
nearest neighbor hopping \cite{kin} in the $x$-direction, i.~e. 
\begin{equation}
\langle k_x \rangle = -t \sum_{\sigma} \langle c^\dagger_{i\,\sigma}c_{i+x\,\sigma}+c^\dagger_{i+x\,\sigma}c_{i\,\sigma} \rangle.
\end{equation}

Since we are only interested in the temperature
region $T \gtrsim T_c$, one can safely assume that the fluctuations of the
phase $\varphi_i$ are predominantly determined by a classical $XY$
free energy 
\cite{em.ki.95,ca.ki.99},
\begin{equation}
F\left[\varphi_i\right] = - J \sum_{\langle ij\rangle}
\cos\left(\varphi_i-\varphi_j\right)\ .
\label{two}
\end{equation}
Our physical picture here is that the XY-action arises from integrating out the shorter 
wavelength fermion degrees of freedom, including those responsible for the 
formation of the local pair amplitude and the internal $d_{x^2-y^2}$ 
structure of the pair. Thus, the \emph{scale} of the XY-lattice spacing is 
actually set by the pair coherence length $\xi_0$. In our work, we have chosen $\Delta$ so that 
$\xi_0 \sim  \frac{v_F}{\pi \Delta} \sim 1$. In this case, the phase configurations 
${\varphi_i}$ calculation can be carried out on the same $L \times L$ $(L=32)$ 
lattice that is used for the diagonalization of the Hamiltonian.
This allows the Kosterlitz-Thouless phase correlation length $\xi$ to grow
over a sufficient range as $T$ approaches $T_{KT}$ and minimizes
finite-size effects.
Thus, we are always in the limit 
where the phase correlation length $\xi$ is larger than the Cooper pair 
size $\xi_0$, when the temperature T is below the mean-field critical
temperature $T_c^{MF}$.

In principle, also the coupling energy $J$ can be considered as arising from
integrating out the high-energy degrees of freedom of the underlying microscopic system. 
Here, we will proceed phenomenologically, neglecting the temperature dependence of $J$ and
simply use it to set the Kosterlitz-Thouless transition temperature
$T_{KT}$ equal to some fraction of $T^{MF}_c$. Specifically, for the
present calculations we will set $T_{KT}\simeq \frac{1}{4} T^{MF}_c$.
 This choice is motivated by the recent scanning-tunneling results
 in $Bi_2Sr_2CuO_{6+\delta}$, where $T_c \simeq 10K$
and the pseudogap regime extends to about 50K, which we take as
$T_c^{MF}$.

In a previous paper \cite{ec.sc.02} we have presented a detailed numerical solution of the 2D-BCS like
Hamiltonian of Eq.~\ref{one} with a $d$-wave gap and phase fluctuations. This is a minimal model but, 
nevertheless, contains the key ideas of the cuprate phase fluctuation scenario:
that is, a $d$-wave BCS gap amplitude forms below a mean-field temperature
$T^{MF}_c$, but phase fluctuation suppress the actual transition
to a considerably lower temperature $T_c$. In the 
intermediate temperature
regime between $T_c^{MF}$ and $T_c$, the phase fluctuations of the gap give
rise to pseudogap phenomena. Comparison of these results
with recent scanning tunneling spectra of Bi-based high-$T_c$ cuprates supports the idea
that the pseudogap behavior observed in these experiments can be understood
as arising from phase fluctuations \cite{ec.sc.02}.

In the present calculations, where we assume a BCS temperature dependence of 
the pairing gap $\Delta(T)$, we have therefore set $\Delta(T=0)=1.0t$ corresponding
to $T^{MF}_c \simeq 0.42 t$ and selected $J$ so that $T_{KT}=0.1t$ \cite{fe.fe.86n}.
The condition that  $\xi>\xi_0$ is thus always
fulfilled, if we are not too close to $T^{MF}_c$.
The calculation of the kinetic energy for an $L\times L$ $(L=32)$ periodic
lattice now proceeds as follows \cite{ref1,da.yu.98n}: a set of phases $\{\varphi_i\}$ is
generated by a Monte Carlo importance sampling procedure, in which the
probability of a given configuration is proportional to $\exp
(-F[\varphi_i]/T)$ with $F$ given by Eq.~(\ref{two}).  With $\{\varphi_i\}$
given, the Hamiltonian of Eq.~(\ref{one}) is diagonalized and the
 kinetic energy $E_{kin}(T,\{\varphi_i\}) = \langle k_x \rangle_{\{\varphi_i\}}$ is extracted.
Further Monte-Carlo $\{\varphi_i\}$ configurations are generated and an
average kinetic energy $E_{kin}(T)= \langle 
k_x \rangle$, at a given temperature, is determined.

\begin{figure}[t]
\begin{center}
\epsfig{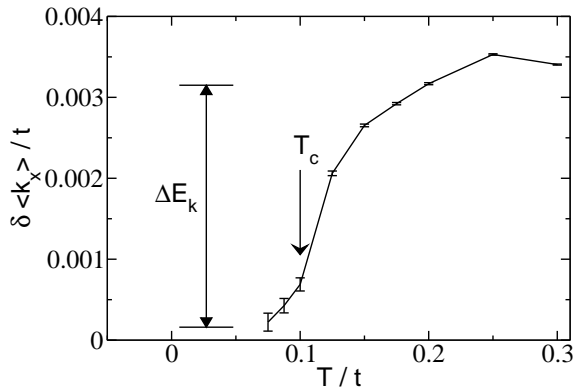}
\end{center}
\caption[]{Kinetic energy contribution from phase fluctuations $\delta\langle k_x \rangle
=\langle k_x \rangle_{PP}-\langle k_x \rangle_{BCS}$. One can clearly see the sharp decrease
of the kinetic energy near the Kosterlitz-Thouless transition at $T=0.1 t \equiv T_c$.
$\Delta E_k$ gives a estimate of the kinetic condensation energy.}
\label{delta_ekin}
\end{figure}

Fig.~\ref{ekin} displays the kinetic energy $\langle k_x \rangle$ as a function of temperature 
for non-interacting tight-binding electrons, for BCS electrons, and for our
phase-fluctuation model, respectively.
We can clearly see that pairing, as expected, produces an overall
increase of  kinetic energy
(indicated as vertical arrows) with respect to the free-electron case.
We observe that in the phase-fluctuation model the kinetic energy is further increased (small
vertical arrows) due to the incoherent motion of the paired electrons.
The kinetic energy is a smoothly decreasing
function of temperature for $T\rightarrow 0$.
This is expected  
from the fact that, at high temperature, more electrons are transferred
to higher kinetic energies, and is
in agreement with the experimental results
\cite{mo.pr.02,sa.lo.01u}. 
What we are especially interested in, is the rather pronounced change 
(magnified in Fig.~\ref{delta_ekin} by using a different scale
for the kinetic energy)
near $T_c \equiv T_{KT}$, 
where the kinetic energy of our 
phase-fluctuation model rather suddenly reduces to the BCS value. 
This sudden deviation from the $T \gtrsim T_c$ behavior is also
  obtained in experiments,
which show 
a kink in the temperature dependence of the
low-frequency spectral weight $W_{low}+W_D$ at $T_c$ \cite{mo.pr.02}.

This pronounced change
of in-plane kinetic energy can be better observed
 in Fig.~\ref{delta_ekin}, where we plot 
the difference between the BCS kinetic energy and the kinetic anergy of our phase-fluctuation model
$\delta\langle k_x \rangle=\langle k_x \rangle_{PP}-\langle k_x \rangle_{BCS}$.
As discussed above, this 
reduction is due to the onset of phase coherence of the Cooper pairs 
below the \sc transition temperature
$T_c \equiv T_{KT}$.
This is signaled by
 the appearance of sharp coherence peaks in the single-particle
 spectral function
upon developing long-range phase coherence~\cite{ec.sc.02}.
The corresponding result for the density of states $N(\omega)$ displaying these coherence peaks is shown in Fig.~\ref{n_omega}.

Notice that this argument for the reduction of kinetic energy
at \tc due to a phase ordering transition
 is quite
robust. For example, we expect it to be valid (and actually to be
stronger) in a true three-dimensional system. As a matter of fact, it
has been argued~\cite{em.ki.95,ca.ki.99}
 that even small interplane couplings play an important
role due to the  infinite-order nature of the
the KT transition.

In order to get a rough estimate of the kinetic condensation energy, we
calculate the reduction in kinetic energy near $T_c$, i.~e.
\begin{equation}
 \Delta E_k = -\frac{2}{e^2 \pi}\int_0^{\infty} ( Re\; \Tilde{\sigma}_{xx}^{sc}(\omega)- 
Re\; \Tilde{\sigma}_{xx}^{n}(\omega))\; d \omega,
\end{equation}
as indicated by the energy change $ \Delta E_k$ in figure \ref{delta_ekin}.
Assuming that $t \simeq 250\, meV$, we get a condensation energy estimate of 
$1.5\,meV $ per Copper site, which is in order of magnitude agreement
with the experimental results (again assuming a one-band TB analysis).

Up to now, to refrain from further approximations, we have set the chemical
potential $\mu$ equal to zero and have only considered nearest-neighbor hopping.
We have checked to some extent, how robust these results are with respect to finite doping
($\langle n \rangle \approx 0.9$) and the inclusion of a next-nearest neighbor-hopping term $t^\prime$
in our Hamiltonian Eq.~\ref{one}. 
Notice that in this case $E_k$ is no longer proportional to the kinetic energy. 
For $t^\prime \lesssim 0.3 t$, our results for
the sum rule \emph{violation} are reduced only by about $20 \% - 30 \%$.

\section{Summary}

In conclusion, we have shown that
the recently observed \emph{violation} of
the low-frequency optical sum rule in the \sc state,
 associated with a
reduction of kinetic energy, can be related to the role of phase fluctuations. 
The decrease in kinetic energy is due to the sharpening of the quasiparticle peaks
close to the superconducting transition at $T_c \equiv T_{KT}$, where the phase
correlation length $\xi$ diverges.
We suggest that this sum rule violation 
should also appear in other superconductors with low charge carrier
density (phase stiffness) such as the organic superconductors.

\section*{Acknowledgments}

We would like to acknowledge useful discussions and comments by S.~A.~Kivelson 
and D.~J.~Scalapino.
This work was supported by the DFG under Grant No.~Ha 1537/16-2 and by a  
 Heisenberg fellowship (AR 324/3-1), by the Bavaria California Technology Center (BaCaTeC),
the  KONWHIR projects OOPCV and CUHE.
The calculations were carried out at the high-performance computing centers 
HLRS (Stuttgart) and LRZ (M\"unchen).


\begin{thebibliography}{10}

\bibitem{em.ki.95}
V.~J. Emery and S.~A. Kivelson, Nature (London) {\bf 374},  434  (1995).

\bibitem{ra.tr.92}
M. Randeria, N. Trivedi, A. Moreo, and R.~T. Scalettar, Phys. Rev. Lett. {\bf
  69},  2001  (1992).

\bibitem{fr.mi.98}
M. Franz and A.~J. Millis, Phys. Rev. B {\bf 58},  14572  (1998).

\bibitem{kw.do.99}
H.-J. Kwon and A.~T. Dorsey, Phys. Rev. B {\bf 59},  6438  (1999).

\bibitem{herb.02}
I.~F. Herbut, Phys. Rev. Lett. {\bf 88},  047006  (2002).

\bibitem{fr.te.01}
M. Franz and Z. Te\v{s}anovi\'{c}, Phys. Rev. Lett. {\bf 87},  257003  (2001).

\bibitem{ec.sc.02}
T. Eckl, D.~J. Scalapino, E. Arrigoni, and W. Hanke, Phys. Rev. B {\bf 66},
  140510(R)  (2002).

\bibitem{ku.fi.01}
M. Kugler, {\O}. Fischer, C. Renner, S. Ono, and Y. Ando, Phys. Rev. Lett. {\bf
  86},  4911  (2001).

\bibitem{co.ma.99}
J. Corson, R. Mallozzi, J. Orenstein, J.~N. Eckstein, and I. Bozovic, Nature
  {\bf 398},  221  (1999).

\bibitem{wa.xu.00}
Y. Wang, Z.~A. Xu, T. Kakeshita, S. Uchida, S. Ono, Y. Ando, and N.~P. Ong,
  Phys. Rev. B {\bf 64},  224519  (2000).

\bibitem{wa.on.02}
Y. Wang, N.~P. Ong, Z.~A. Xu, T. Kakeshita, S. Uchida, D.~A. Bonn, R. Liang,
  and W.~N. Hardy, Phys. Rev. Lett. {\bf 88},  257003  (2002).

\bibitem{ru.al.03u}
F. Rullier-Albenque, H. Alloul, and R. Tourbot, cond-mat/0301596 (unpublished).

\bibitem{mo.pr.02}
H.~J.~A. Molegraaf, C. Presura, D. van~der Marel, P.~H. Kes, and M. Li, Science
  {\bf 295},  2239  (2002).

\bibitem{sa.lo.01u}
A. Santander-Syro, R. Lobo, N. Bontemps, Z. Konstantinovic, Z. Li, and H.
  Raffy, cond-mat/0111539 (unpublished).

\bibitem{tinkham}
M. Tinkham, {\em Introduction to Superconductivity} (McGraw--Hill, New York,
  1975).

\bibitem{ha.sh.80}
W. Hanke and L.~J. Sham, Phys. Rev. B {\bf 21},  4656  (1980).

\bibitem{hirs.02}
Hirsch, Science {\bf 295},  2226  (2002).

\bibitem{no.pe.02u}
M.~R. Norman and C. Pepin, Phys. Rev. B {\bf 66},  100506(R) (2002).

\bibitem{kubo.57}
R. Kubo, J. Phys. Soc. Jpn. {\bf 12},  570  (1957).

\bibitem{ca.ki.99}
E.~W. Carlson, S.~A. Kivelson, V.~J. Emery, and E. Manousakis, Phys. Rev. Lett.
  {\bf 83},  612  (1999).

\bibitem{klei.00}
H. Kleinert, Phys. Rev. Lett. {\bf 84},  286  (2000).

\bibitem{re.re.p.98}
C. Renner, B. Revaz, J.-Y. Genoud, K. Kadowaki, and {\O}. Fischer, Phys. Rev.
  Lett. {\bf 80},  149  (1998).

\bibitem{mi.za.99}
N. Miyakawa, J.~F. Zasadzinski, L. Ozyuzer, P. Guptasarma, D.~G. Hinks, C.
  Kendziora, and K.~E. Gray, Phys. Rev. Lett. {\bf 83},  1018  (1999).

\bibitem{kivelson}
The concept of a kinetic-energy driven mechanism for superconductivity is
  discussed in detail in Ref.~\cite{ca.em.02u}.

\bibitem{sc.wh.93}
D.~J. Scalapino, S.~R. White, and S. Zhang, Phys. Rev. B {\bf 47},  7995
  (1993).

\bibitem{delta}
Our mean-field Hamiltonian, Eq.~\ref{one} can be seen as derived by decoupling
  an interaction term of the form $\Delta_{i\,\delta}^\dagger
  \Delta_{i\,\delta}$, which does not couple to the gauge field and, thus, does
  not contribute to the current.

\bibitem{kin}
This term is commonly referred to as ``kinetic energy'' to distinguish it from
  the interacting part of the Hamiltonian.

\bibitem{fe.fe.86n}
$T_{KT} \simeq 0.89 J$ see, for example, J.~F.~Fern\'{a}ndez, M.~F.~Ferreira,
  and J.~Stankiewicz, Phys.~Rev.~B {\bf 34}, 292 (1986).

\bibitem{ref1}
For a more detailed discussion, see: T.~Eckl {\it et al.}~(to be published);
  N.~E.~Bickers and D.~J.~Scalapino, cond-mat/0010480 (unpublished); and
  P.~Monthoux and D.~J.~Scalapino, Phys.~Rev.~B {\bf 65}, 235104 (2002).

\bibitem{da.yu.98n}
See also: E.~Dagotto {\it et al.}, Phys.~Rev.~B {\bf 58}, 6414 (1998); and
  H.~Monien, J.~Low Temp.~Phys.~{\bf126}, 1123 (2002).

\bibitem{ca.em.02u}
E.~W. Carlson, V.~J. Emery, S.~A. Kivelson, and D. Orgad, cond-mat/0206217.
  Review chapter to appear in `The Physics of Conventional and Unconventional
  Superconductors' ed. by K. H. Bennemann and J. B. Ketterson (Springer-Verlag,
  Berlin, in press).

\end{thebibliography}

\end{document}